\documentclass[prb,preprint,showpacs,preprintnumbers,amsmath,amssymb]{revtex4}

\usepackage{graphicx}
\usepackage{amsmath}
\usepackage{makeidx} 
\newcommand{\comment}[1]{} 

\begin{document}

\title{
	Resolution-dependent mechanisms for bimodal switching-time distributions in simulated Fe nanopillars
      }

\author{
	S.H. Thompson,$^{1,2,3}$ G. Brown,$^{1,4}$ A. Kuhnle,$^{1,2}$ P.A. Rikvold,$^{1,3,5}$ M.A. Novotny$^{6,7}$
	}

\affiliation{
	$^1$Department of Physics,	\\
	Florida State University, Tallahassee, FL 32306-4350, USA\\
	$^2$Department of Scientific Computing, \\
	Florida State University, Tallahassee, FL 32306-4120, USA\\
	$^3$Center for Materials Research and Technology, \\
	Florida State University, Tallahassee, FL 32306-4350, USA\\
	$^4$Center for Nanophase Materials Science,\\
	Oak Ridge, TN 37831-6164, USA \\
	$^5$National High Magnetic Field Laboratory, \\
	Tallahassee, FL 32310-3706, USA\\
	$^6$Department of Physics and Astronomy, \\ 
	Mississippi State University, Mississippi State, MS 39762, USA\\
	$^7$HPC$\,^2$ Center for Computational Sciences,\\
	Mississippi State University, Mississippi State, MS 39762, USA\\
}
\date{\today}

\begin{abstract}
We study the magnetization-switching statistics following reversal of the applied field for three separate computational models representing the same physical system, an iron nanopillar.
The primary difference between the models is the resolution of the computational lattice and, consequently, the intrinsic parameters that must be rescaled to retain similarity to the physical system.
Considering the first-passage time to zero for the magnetization component in the longitudinal (easy-axis) direction, we look for applied fields that result in bimodal distributions of this time for each system and compare the results to the experimental system.
We observe that the relevant fluctuations leading to bimodal distributions are different for each lattice resolution and result in magnetization-switching behavior that is unique to each computational model. 
Correct model resolution is thus essential for obtaining reliable numerical results for the system dynamics.
\end{abstract}

	
\maketitle

\section{Introduction}
\label{Intro}
The constant growth of the performance of semiconductors, typically characterized by Moore's law, is also witnessed in the progress of data storage.~\cite{McDaniel:ulttamr}
For example, current candidate materials for magnetic storage devices are now poised to surpass an areal density of one terabit/cm$^2$, an increase of several orders of magnitude in just ten years.~\cite{McDaniel:ulttamr}
At this density, the size of the recording bit approaches a limit constrained by superparamagnetism. 
This issue is additionally complicated by the requirement that bits should maintain $95\%$ of their magnetization over a period of ten years to meet the industry standard.
In addition, sub-nanosecond magnetization-switching times are necessary to achieve suitable rates for read/write operations.~\cite{McDaniel:ulttamr}
Storage bits should consequently have the property of a single characteristic switching time, which ensures a predictable response as the read/write head of the device passes over the recording medium.
Along with meeting the numerous manufacturing challenges facing the implementation of these requirements, a comprehensive characterization of the magnetization switching of the constituent small, magnetic particles is needed.

In this paper, we study the switching statistics of models of elongated iron nanopillars, using several different resolutions of the computational lattice to model the same physical pillar.
The high aspect ratio of these systems introduces a shape-induced anisotropy that assists in raising the coercivity of the particle and reducing unwanted thermally-activated switching during the long-term storage of the bit information. 
Here, we are particularly interested in bimodal distributions of the switching time arising in various regimes of the applied field, which could potentially compromise the reliability of the switching process.

A bimodal distribution was first witnessed in a highly resolved model of an iron nanopillar based on experimental work by von Moln\'{a}r and collaborators.~\cite{Thompson:rmsinoof, Kent:goharnsmwcvdastm, McCord:ddomduastm}
Preliminary results for this model indicated a bimodal distribution of switching times near the minimum switching field $H_\mathrm{sw}$ for an obliquely aligned applied field.~\cite{Thompson:rmsinoof}
However, as a consequence of the complexity of the numerical model, it has proved difficult to adequately describe the underlying mechanism responsible for the observed switching-time distribution.
To obtain a more comprehensive understanding, in this paper we look for bimodal distributions of the switching time for lower-resolution models of the same system.
Our results show that the lattice resolution is very significant in that it determines the degree to which fluctuations in the model affect the numerically observed behaviors.

The rest of this paper is organized as follows.
In Sec.~\ref{MaND} we briefly discuss our computational models and describe the numerical procedure that is used in all the simulations.
The nanopillar is modeled at three different resolutions of the computational lattice: high, medium, and low.
Results from the simulations are presented in Secs.~\ref{HR}, \ref{MR}, and \ref{LR}, respectively.
Finally, we present our conclusion in Sec.~\ref{Con}.

\section{Model and Numerical Details}
\label{MaND}
Our numerical models are motivated by real iron nanopillars fabricated by von Moln\'{a}r and collaborators using scanning-tunneling-microscopy-assisted chemical-vapor deposition.~\cite{Kent:goharnsmwcvdastm, McCord:ddomduastm}
Their switching-time results for an approximately $10\times10\times150$~nm$^3$ nanopillar gave a much lower switching field for applied fields directed close to the easy axis than predicted by the Stoner-Wohlfarth model.~\cite{Li:mriefn, Wirth:tamrinsip, Li:hmoasin, Wirth:famponspa}
This is attributed to nonuniform modes of the magnetization and endcap formation that cannot be explained by a Stoner-Wohlfarth type of coherent-rotation model for the magnetization switching.

The highest-resolution model for their experimental nanopillar has a lattice discretization on the order of the physical exchange length $l_e$.
Although this discretization provides the most realistic behavior, the simulation time prevents a statistical description of the switching-time distribution for more than a few values of the applied field.
For the medium-resolution model the lattice is discretized to the width of the pillar, spanning several $l_e$ for each computational cell.
This model, along with the lowest-resolution, single-spin model, allow for a more thorough investigation of switching statistics over a larger region of the applied-field space.
We consequently use results from the highest-resolution model to explore internal magnetization dynamics and compare the resulting switching statistics to those of the lower-resolution models.

All three computational models in this paper employ the stochastic, partial differential Landau-Lifshitz-Gilbert (LLG) equation, 
\begin{equation}
\frac{{\it{d}}{\vec{m}(\vec{r_i})}}{\it{dt}} = -\frac{\gamma_0}{1+\alpha^2}\left(\vec{m}(\vec{r_i})\times 
\left[\vec{H}(\vec{r_i})+\frac{\alpha}{m_\mathrm{s}}\vec{m}(\vec{r_i})\times \vec{H}(\vec{r_i})\right]\right),
\label{eq:M}
\end{equation}
as the method for determining the time evolution of the magnetization for each computational lattice site.~\cite{Brown:IEEE, Aharoni:itf, Brown:m, Brown:lsotamrinp}
Here, $\vec{m}(\vec{r_i})$ is the magnetization at the $i$th lattice site in the presence of the local field $\vec{H}(\vec{r_i})$.  
This local field is a linear combination of individual fields,
\begin{equation}
\vec{H}=\vec{H_\mathrm{Z}}+\vec{H_\mathrm{E}}+\vec{H_\mathrm{D}},
\end{equation}  
with $\vec{H_\mathrm{Z}}$ the applied (Zeeman) field, $\vec{H_\mathrm{E}}$ the exchange field, and $\vec{H_\mathrm{D}}$ the magnetostatic (demagnetizing) field.
Also present is a stochastic thermal field $\vec{H_\mathrm{T}}$, which is treated differently by the integration routine.~\cite{Brown:lsotamrinp}
The thermal field has zero mean and variance given by the fluctuation-dissipation relation,
\begin{equation}
{\langle}{H_{\beta}^\mathrm{T}}({\vec{r_i}},t){H_{\gamma}^\mathrm{T}}(\vec{r_j},t'){\rangle} = 
	\frac{2{\alpha}{k_\mathrm{B}}T}{{\gamma_0}m_\mathrm{s}{V}}\delta_{ij}\delta_{\beta\gamma}\delta(t-t'),
\end{equation}
where $k_\mathrm{B}$ is Boltzmann's constant, $V$ is the volume of an individual computational cell, $T$ is the absolute temperature, $\delta_{ij}$ and $\delta_{\beta\gamma}$ are Kronecker deltas over the lattice sites $i, j$ and directions $\beta, \gamma$, respectively, and $\delta(t-t')$ is a Dirac delta function of the time difference, $t-t'$.
This equation implies that the magnitude of the thermal field scales linearly with the square root of the temperature.
All simulations reported in this paper were performed at $20.2$~K unless specified otherwise.
Constants in the LLG include the electronic gyromagnetic ratio $\gamma_0=1.76 \times 10^7$~Hz/Oe, the saturation magnetization of bulk iron $m_\mathrm{s} = 1700$~$\mathrm{emu/cm^3}$, and a phenomenological damping parameter $\alpha = 0.1$.~\cite{Thompson:rmsinoof, Brown:lsotamrinp}

All three models discussed here use the same field-reversal protocol for the simulation time $-0.125$~ns~$<t<0$~ns.
At $t=-0.125$~ns, the computational pillar is uniformly magnetized along the easy axis in the positive $z$-direction and is subjected to an applied field that is initially anti-parallel to its final direction, which is its direction for $t \ge 0$~ns.
Specifically, the value of the applied field is changed sinusoidally during the field-reversal period to its final value, i.e., $\vec{H}(t) = \vec{H_0}\cos(\pi{t}/0.125~\mathrm{ns})$ with $t\in[-0.125~\mathrm{ns}, 0]$.
For $t > 0$, $\vec{H}(t) = \vec{H_0}$ and remains constant with a negative $z$-component.
This protocol is used to keep the Zeeman energy from excessively exciting the system.
After the completion of the field-reversal protocol at $t=0$, measurement of the switching time begins and only depends upon the value of the $z$-component of the total magnetization $M_z = (1/{(N_{s}m_s)})\sum_{i} m_{z}(\vec{r_i})$, with $N_s$ the total number of computational spins.
Consequently, the switching time is defined as the first-passage time (FPT) to $M_z \leq 0$, starting from $M_z > 0$.

\begin{figure}[tb]
\centering
$\begin{array}{c@{\hspace{0.0truecm}}c}
\includegraphics[scale=0.40]{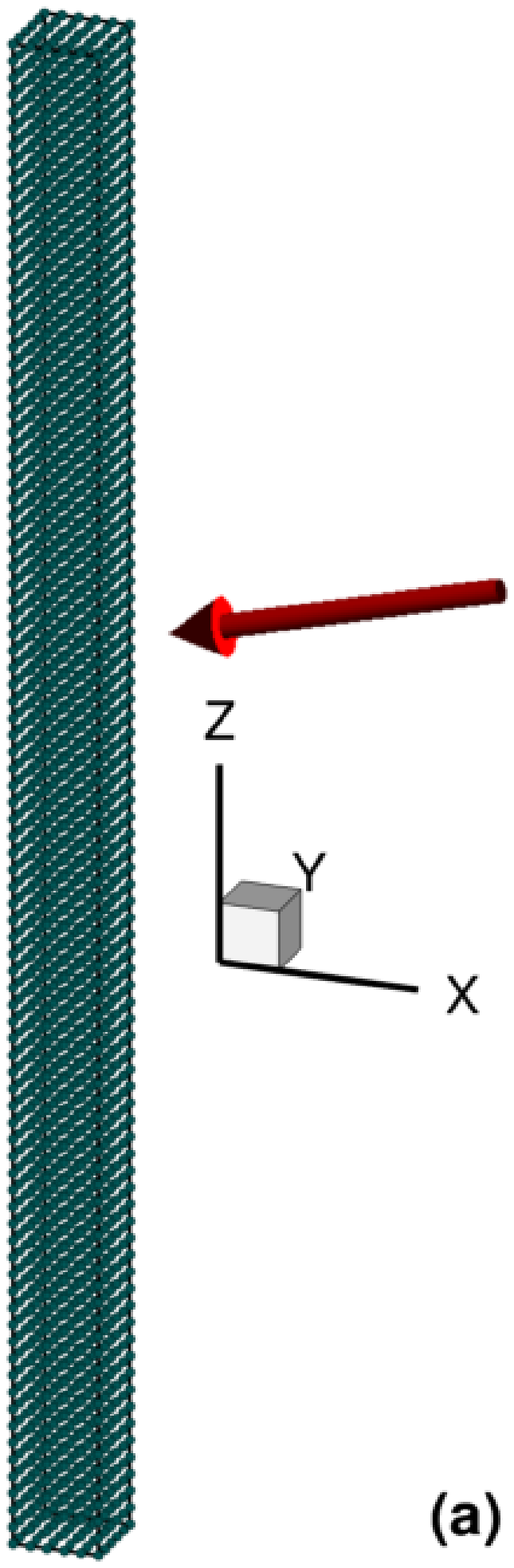}
\includegraphics[scale=0.40]{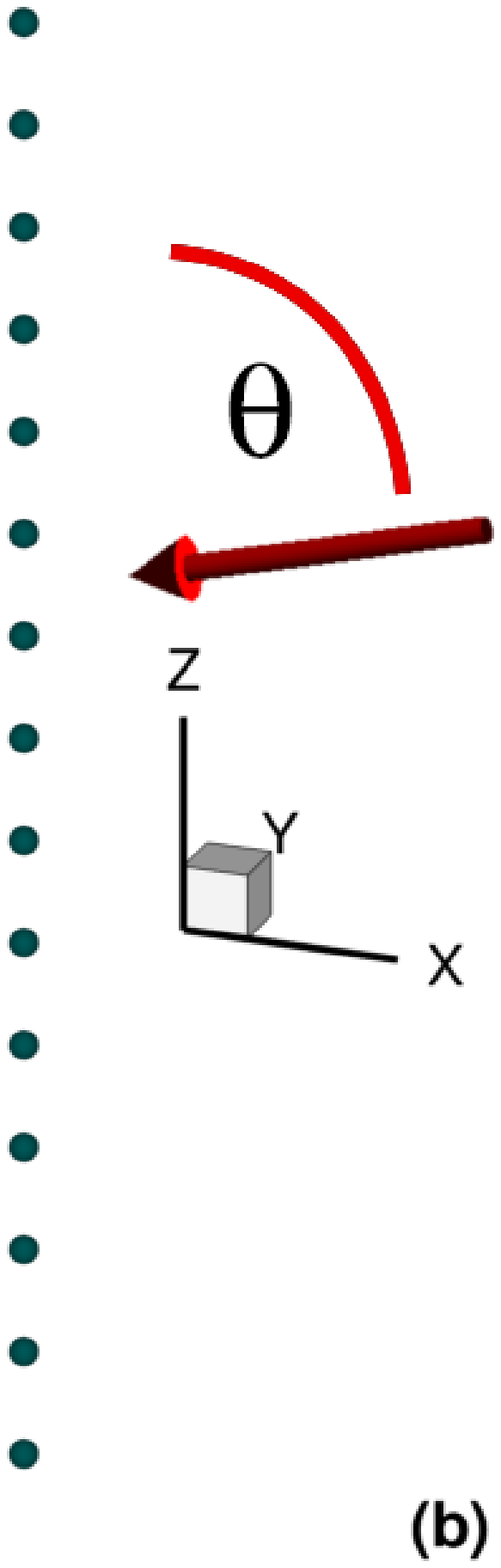}
\includegraphics[scale=0.40]{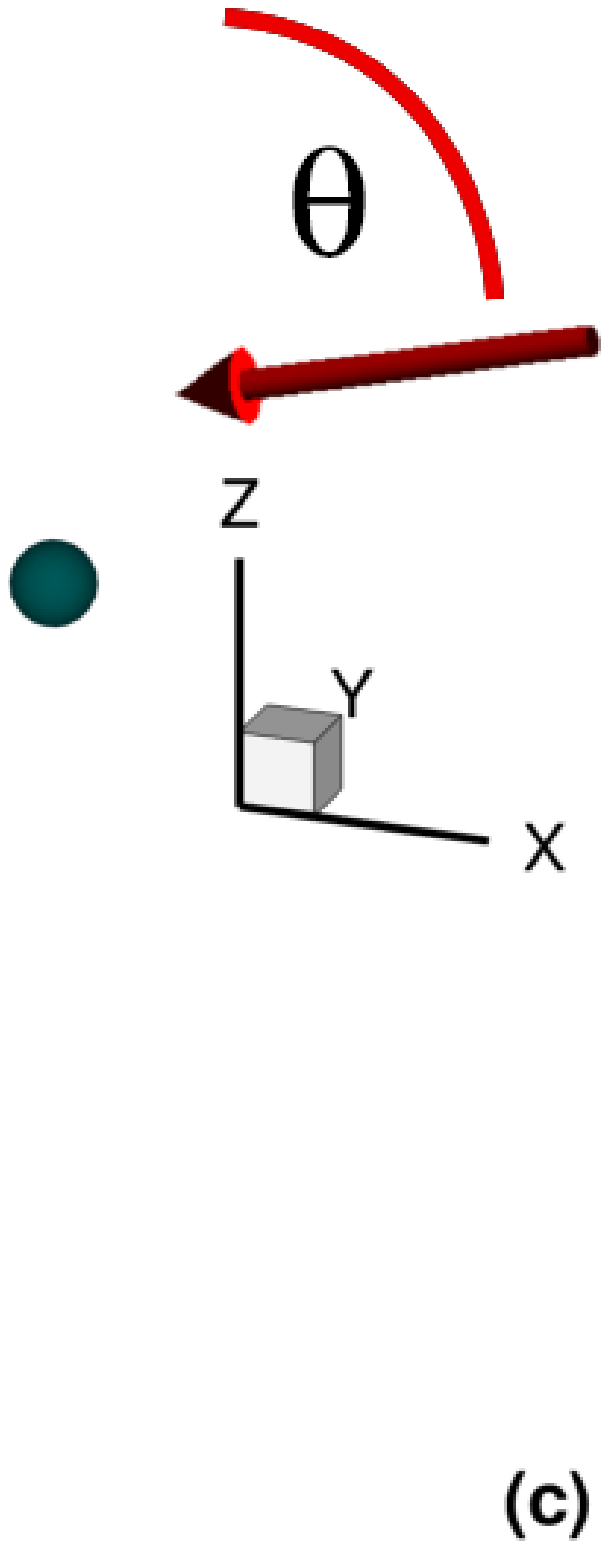}
\end{array}$
\caption[]
{
(Color online) Lattice resolution geometries for the (a) high-resolution model, (b) medium-resolution model, and (c) single-spin model.
Each computational cell is centered on a sphere in the figure.
The final orientation for the applied field of the high-resolution model is included as a bold arrow (red online) in the figures.
Other orientations of the applied field are also used in this study for the medium and low-resolution models, with the $y$-component of the applied field always zero.
}
\label{fig:lattice}
\end{figure}

Figure~\ref{fig:lattice} displays the computational cell geometry for the three models discussed in this paper: (a) the high-resolution $6\times6\times90$ cells, (b) the medium-resolution $1\times1\times15$ cells, and (c) the single-spin model.
An example of the final direction of $\vec{H_\mathrm{Z}}$ is also shown beside each lattice as a bold arrow.
For the high-resolution model, the switching statistics are collected for only one value of $\vec{H_\mathrm{Z}}$ due to the computational time required to gather a statistically significant amount of data.
The switching statistics of the low and medium-resolution models, however, are studied as functions of $H_{\mathrm{Z}x}$ and $H_{\mathrm{Z}z}$.

Switching near the coercive field involves spins at the ends of the pillar for both the high and medium-resolution models.
For the high-resolution model, this is a result of the initial formation of endcaps, regions of large curl at the ends of the pillar that lower the free energy through pole avoidance.
Since end spins of the medium-resolution model have only one nearest neighbor, they can have larger changes in orientation for a given energy cost compared to the internally located spins.
The center of the pillar between the two ends remains mostly uniform while in the metastable state for both of these models, except for small thermal fluctuations and propagating low-amplitude spin waves.
Eventually, the collective, random thermal fluctuations carry one or both ends out of the metastable free-energy well and allow magnetization switching to occur.
The macrospin model, however, only has to exit from a simple two-dimensional metastable free-energy well and switches via a precession that dissipates energy through the damping term in the LLG equation.

\section{High Resolution}
\label{HR}
First, we describe results for this nanopillar system with a three-dimensional, high-resolution, $6\times6\times90$ cells computational model that possesses a lattice discretization, $\Delta{r_i}$, which is smaller than the exchange length of $l_e = 2.6$~nm for the real system.~\cite{Thompson:rmsinoof}
Since the magnetization of the real system does not change appreciably across $\Delta{r_i}<{l_e}$, this resolution provides the most realistic internal magnetization dynamics of all of our computational models.
A further decrease of $\Delta{r_i}$ should not provide significantly increased accuracy.

\begin{figure}[tb]
\centering\includegraphics[scale=0.40]{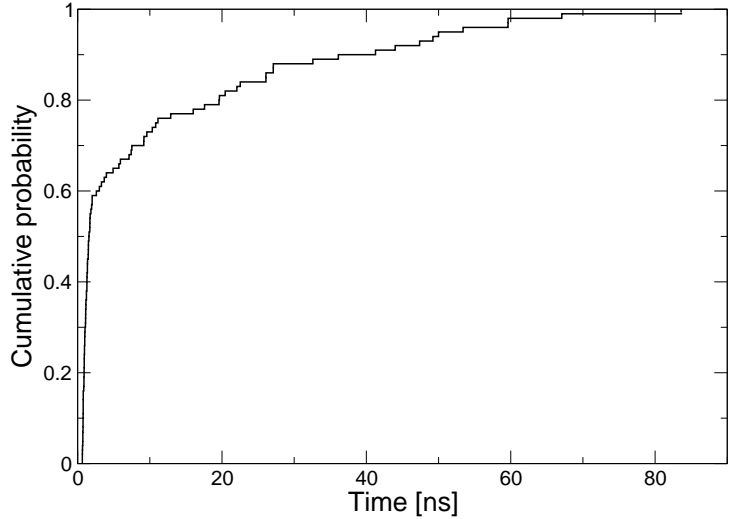}
\caption[]
{
Cumulative switching-time distribution at $20.2$~K for the highly resolved model for the applied field $\vec{H_\mathrm{Z}} = 3260$~Oe at $75^\circ$ with respect to the long axis of the pillar.
The bimodal behavior seen in this distribution is the result of a magnetization-switching process in which the endcap may or may not configure itself into a long-lived metastable configuration.
}
\label{fig:CD_Full}
\end{figure}

When this model is subjected to a near-coercive applied field, $\vec{H_\mathrm{Z}} = 3260$~Oe at $75^\circ$ with respect to the long axis of the pillar, a bimodal distribution of FPTs is obtained.~\cite{Thompson:rmsinoof}
This distribution, shown in Fig.~\ref{fig:CD_Full} as a cumulative distribution for $100$ trials, is divided into two groups based solely on the observed distribution: a fast mode (switching times $<2.5$~ns) and a slow mode (switching times $>2.5$~ns).
Both modes are fitted well by the delayed exponential, $f(t) = \eta(t-t_0)(1/\tau)\exp(-(t-t_{0})/\tau)$, where $\eta$ is the Heaviside step function and $t_{0} = \min[\min\{t_i\}, \langle{t}\rangle - \sigma_{t}]$, with $\tau=0.5$~ns for the fast mode and $\tau=21.7$~ns for the slow mode.
A detailed analysis of the results of the high-resolution model will be presented elsewhere.~\cite{Thompson:inPress}
Here we provide only those details needed to compare and contrast with the medium and low-resolution models.

We found that the bimodal switching-time distribution is the result of multiple switching paths through a high-dimensional free-energy landscape, each path with a single characteristic switching time that the system chooses with almost equal probability for the applied field used in this study.
Measurements of the total energy $E$ during the simulations were nearly constant during the switching process, indicating that changes in the free energy $F=E-TS$ mostly come from changes in the entropy $S$.
Fluctuations of the coarse-grained spins allow trajectories to enter a region of the free-energy space influenced by a local minimum that requires a large decrease in $S$ to exit, resulting in the slow mode.
If the fluctuations do not cause the pillar to fall into this metastable free-energy well, a fast mode is observed for which the trajectories follow a free-energy path with almost constant $F$, corresponding to an entropy change that is relatively small. 
The difference in the magnetization configurations of the endcaps between the two modes is subtle, however, and it has proved difficult to identify the mode solely by analyzing the endcaps.
For example, the volumes of the endcaps are nearly the same for both the fast and slow modes and agree with the experimental fit of the activation volume yielding $v_A \approx 270$~nm$^3$.~\cite{Wirth:mbonsip}
Further characterization of the endcap configuration did not reveal differences that would indicate if switching occurred via a fast or a slow mode. 

Since switching is equally likely to initiate at either endcap, fast modes occur if either one or both endcaps do not pass through the long-lived configuration.
However, it is necessary for both endcaps to explore the longer-lived metastable configuration in order to qualify as a slow mode.
For the high-resolution pillar, we also find that the fast-mode statistics are not dependent on the number of endcaps that switch (one vs two).
However, this detail is central to the explanation of the bimodal distribution seen in the medium-resolution model discussed in Sec.~\ref{MR}.

\section{Medium Resolution}
\label{MR}
The medium-resolution model is a one-dimensional stack of spins, with a lattice resolution of $1\times1\times15$ cells.
This model does not allow for variation of the magnetization in the transverse-to-easy-axis direction, but permits non-uniform magnetization along the spin chain.
Due to the considerably smaller computational time required by this and the lowest-resolution models, it is feasible to sample the switching statistics for a much larger region in the applied-field space, $\vec{H_\mathrm{Z}}$.

\begin{figure}[tb]
\centering\includegraphics[scale=0.40]{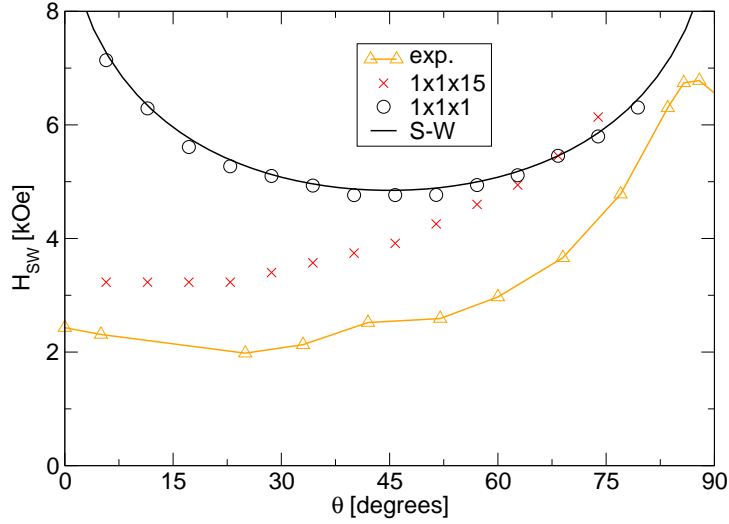}
\caption[]
{
(Color online) Minimum switching field $H_\mathrm{sw}$ at $30.3$~K as a function of the angle of the applied field.
The $1\times1\times15$ medium-resolution, stack-of-spins model (crosses, red online) deviates from the macrospin model (circles, black) at $\theta < 60^\circ$ in agreement with experimental observations (triangles, orange online).~\cite{Li:mriefn}
The $1\times1\times1$ macrospin model displays a Stoner-Wohlfarth type of behavior.
Analytical results for a Stoner-Wohlfarth spin are shown as a black curve.
For the numerical data, $H_\mathrm{sw}$ is defined as the field that causes $50\%$ of the trials to switch for a waiting time of $3.34$~ns.
Error bars for all results are on the order of the symbol size.
}
\label{fig:HswTheta}
\end{figure}

\begin{figure}[tb]
\centering\includegraphics[scale=0.40]{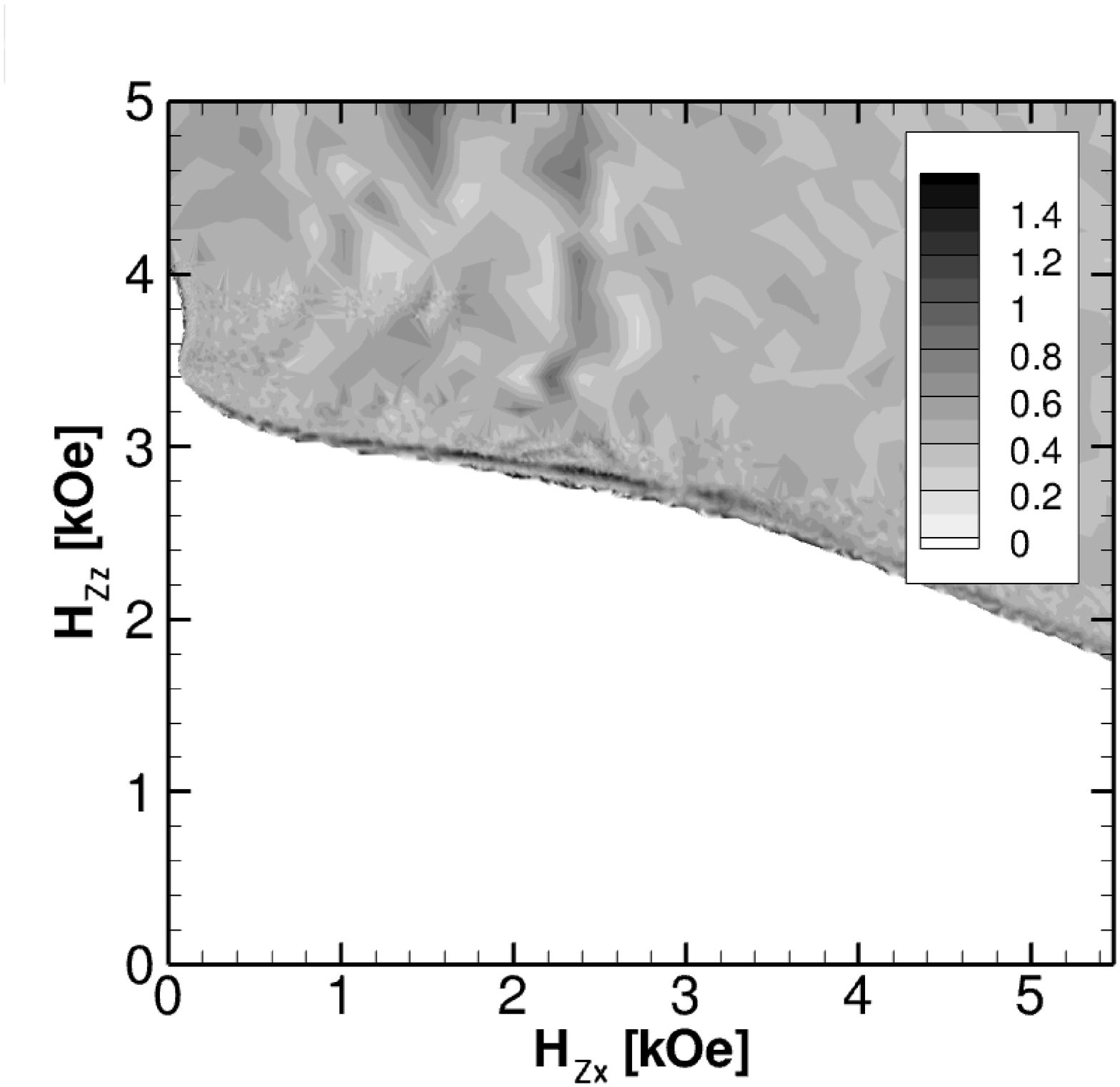}
\caption[]
{
Applied-field space at $20.2$~K for the medium-resolution $1\times1\times15$ model with a waiting time of $3.34$~ns. 
Larger values of the variable ($C>1$) may indicate the existence of more than one characteristic time in the switching-time distribution.
Bimodal distributions are seen just above the coercive edge in this figure as a ridge that almost extends across the entire length of the $H_{\mathrm{Z}x}$-axis.
The empty white area below this ridge denotes the region where switching times are larger than the waiting time.
}
\label{fig:switchspace_med}
\end{figure}

Figure~\ref{fig:HswTheta} reveals the minimum switching field $H_\mathrm{sw}$ as a function of $\theta$, the angle of the applied field with respect to the easy axis of the pillar, for $T = 30.3$~K ($T$ chosen to match the experimental conditions).
This figure is generated from data that have $100$ trials per applied-field value, with $H_\mathrm{sw}$ defined as the field that causes $50\%$ of the trials to switch for a waiting time of $3.34$~ns.
Qualitatively similar results are also found for $H_\mathrm{sw}$ using different waiting times.
The experimental results are also shown in Fig.~\ref{fig:HswTheta} for comparison, and show good agreement with the medium-resolution model. 
In particular, both the $1\times1\times15$ model and the experimental results show a deviation from a Stoner-Wohlfarth uniform-rotation type of behavior for $\theta < 60^\circ$.~\cite{Li:mriefn}
Analytical results for a Stoner-Wohlfarth spin are also presented in the figure and match the $\theta$-dependence of $H_\mathrm{sw}$ for the lowest-resolution model discussed in Sec.~\ref{LR}.~\cite{Stoner:amomhiha}
The dependence of the minimum switching field on the orientation of the applied field is also revealed by the location of the coercive edge in the applied-field space of Fig.~\ref{fig:switchspace_med}.

We begin searching for bimodal distributions for this system by plotting the variable $C$ in Fig.~\ref{fig:switchspace_med}, defined as
\begin{equation}
C = \frac{\sigma}{\mathrm{mean}\{t_i\}-\mathrm{min}\{t_i\}},
\label{eq:C}
\end{equation}  
where $\sigma = \sqrt {(1/(N-1)) \sum_{i}(t_i-\langle{t}\rangle)^2}$ is the standard deviation of the observed switching times $t_i$ over a sample of $N=100$ trials, $\mathrm{mean}\{t_i\}$ is the mean value, and $\mathrm{min}\{t_i\}$ is the minimum value of the sample.
Values of $C$ that are greater than unity may indicate the existence of a switching-time distribution with more than one characteristic time.

Bimodal switching-time distributions are seen as a ridge that almost extends across the entire $H_{\mathrm{Z}x}$-axis and is located just above the coercive edge of Fig.~\ref{fig:switchspace_med}.
As the ridge is crossed, the ratio of faster to slower switching times changes from a larger percentage of slower times just below the ridge, to a larger percentage of faster times just above the ridge.
The larger values of $C$ observed at the coercive edge preceding the bimodal ridge, most easily seen near $\theta = 45^\circ$, are the result of incomplete switching statistics with $N < 5$.
An increase in the maximum waiting time of the simulation should improve the accuracy of $C$ in this region.
It should also be noted that larger values of $C$ seen in the interior of the plot (e.g., near $H_{\mathrm{Z}x} = 2.4$~kOe and $H_{\mathrm{Z}z} = 4.5$~kOe) are caused by applied fields that are large enough to cause switching when $t<0$, during the initial field reversal.
\begin{figure}[tb]
\centering\includegraphics[scale=0.40]{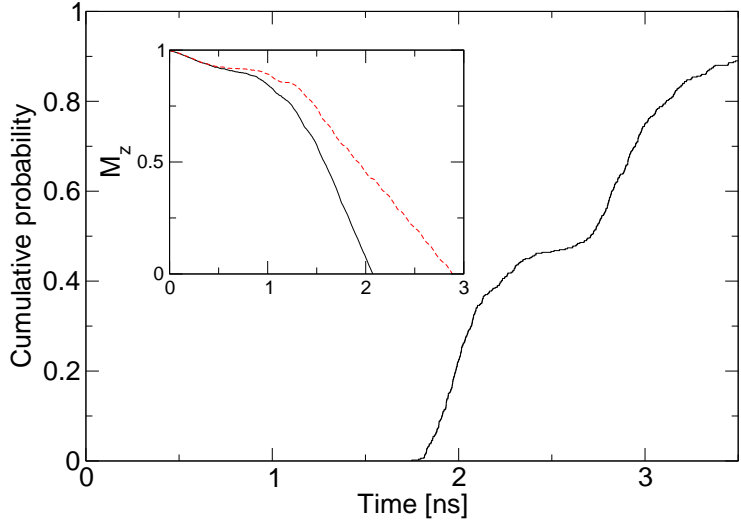}
\caption[]
{
(Color online) Cumulative switching-time distribution for the medium-resolution pillar at $20.2$~K for the applied field $H_{\mathrm{Z}x} = 340$~Oe and $H_{\mathrm{Z}z} = 3.2$~kOe.
$N=10000$ trials were used to generate this figure, with $9043$ returning a switching time below the waiting time of $6.68$~ns.
The $z$-component of the global magnetization for two trials is also shown as an inset.
A faster trial (black, solid) shows a slope that is twice as large as that of the slower trial (dashed, red online), resulting from both endcaps releasing at approximately the same time.
}
\label{fig:CD_1x1x15a}
\end{figure}

The cumulative switching-time distribution for the field $H_{\mathrm{Z}x} = 340$~Oe and $H_{\mathrm{Z}z} = 3.2$~kOe, located close to the center of the bimodal ridge,  is presented in Fig.~\ref{fig:CD_1x1x15a}.
An interesting feature of this distribution, compared to the highly resolved model, is the clear separation of faster and slower switching times. 
Since nucleation of the domain wall is equally likely to occur at either end of the pillar, two possible scenarios may happen during switching for this model.
For the slow mode, nucleation of the domain wall only happens at one end of the pillar.
As the wall proceeds along the pillar, the dipolar field is lowered for the spin situated at the opposite end of the pillar, preventing that spin from nucleating another domain wall.
Growth toward the stable state of the pillar happens in this case at a rate that is given by the movement of a single domain wall.
On the other hand, the fast mode is the result of nucleation of domain walls at both ends of the pillar at nearly the same time, with a corresponding change in $M_z$ that occurs approximately twice as fast for the faster mode.
This is revealed in the inset of Fig.~\ref{fig:CD_1x1x15a} as the slope of the global magnetization $M_z$ vs time $t$. 
Since either both end spins have to nucleate at nearly the same time, before the dipolar field from the switching region increases the nucleation barrier for the opposite end, or one at a time, a clear separation of observed switching times is seen in Fig.~\ref{fig:CD_1x1x15a}.
This indicates that the ends of the pillar are coupled since switching from either end affects the opposite end.
If this were not the case, one should expect a greater number of switches to occur near $t=2.5$~ns in Fig.~\ref{fig:CD_1x1x15a}.
\begin{figure}[tb]
\centering
$\begin{array}{c@{\hspace{0.0truecm}}c}
\includegraphics[scale=0.33]{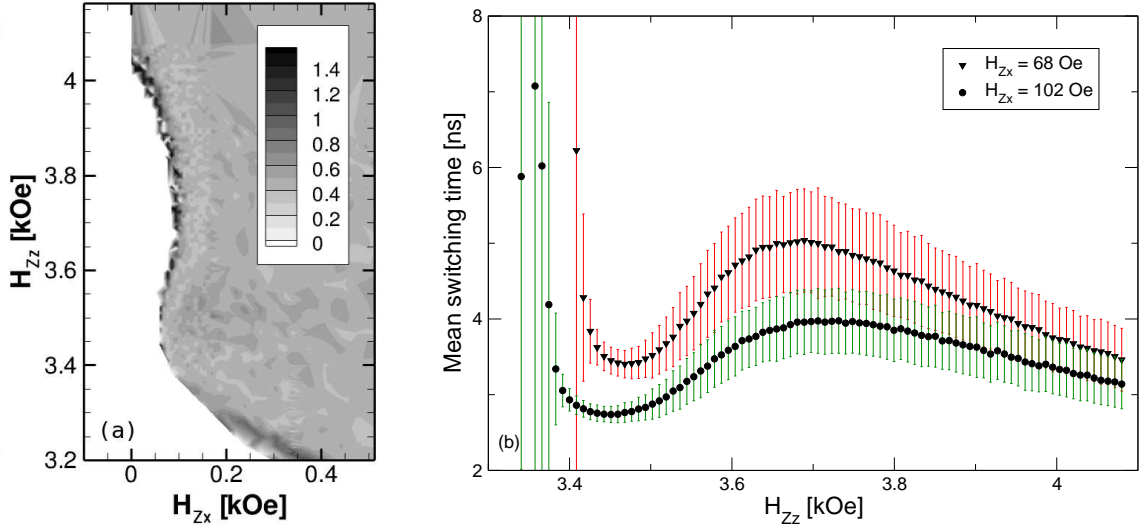}
\includegraphics[scale=0.40]{6b_MedSw}
\end{array}$
\caption[]
{
(Color online) (a) Closer view of the ``reentrant'' region for the medium-resolution $1\times1\times15$ model.  
The white region has switching times greater than the waiting time.
(b) Mean switching times for $H_{\mathrm{Z}x} = 68$~Oe (triangles, upper curve) and $H_{\mathrm{Z}x} = 102$~Oe (circles, lower curve) with the error bars indicating one standard deviation.
Data used for (b) are the result of $1000$ trials per point, with a maximum waiting time of $33.42$~ns.
}
\label{fig:MedClose}
\end{figure}

Another interesting feature of the medium-resolution system is the region of ``reentrant'' behavior, revealed in Fig.~\ref{fig:switchspace_med} as the concave region of the coercive edge near the $H_{\mathrm{Z}x} = 0$~kOe axis.
Figure~\ref{fig:MedClose}(a) provides a magnified view of this feature using $C$ defined in Eq.~(\ref{eq:C}).
As can be seen in Fig.~\ref{fig:MedClose}(b), which plots the mean switching time vs $H_{\mathrm{Z}z}$ for $H_{\mathrm{Z}x} = 68$~Oe and $102$~Oe, the mean switching time is a nonmonotonic function of $H_{\mathrm{Z}z}$ and increases with increasing applied field for $H_{\mathrm{Z}z} = 3.5$~kOe to $H_{\mathrm{Z}z} = 3.7$~kOe.
This behavior is not confined to the region close to the $H_{\mathrm{Z}z}$-axis.
The switching times of the spin chain remain nonmonotonic for $H_{\mathrm{Z}z} \le 3.7$~kOe in the region of $H_{\mathrm{Z}x} < 2.5$~kOe.
\begin{figure}[tb]
\centering\includegraphics[scale=0.60]{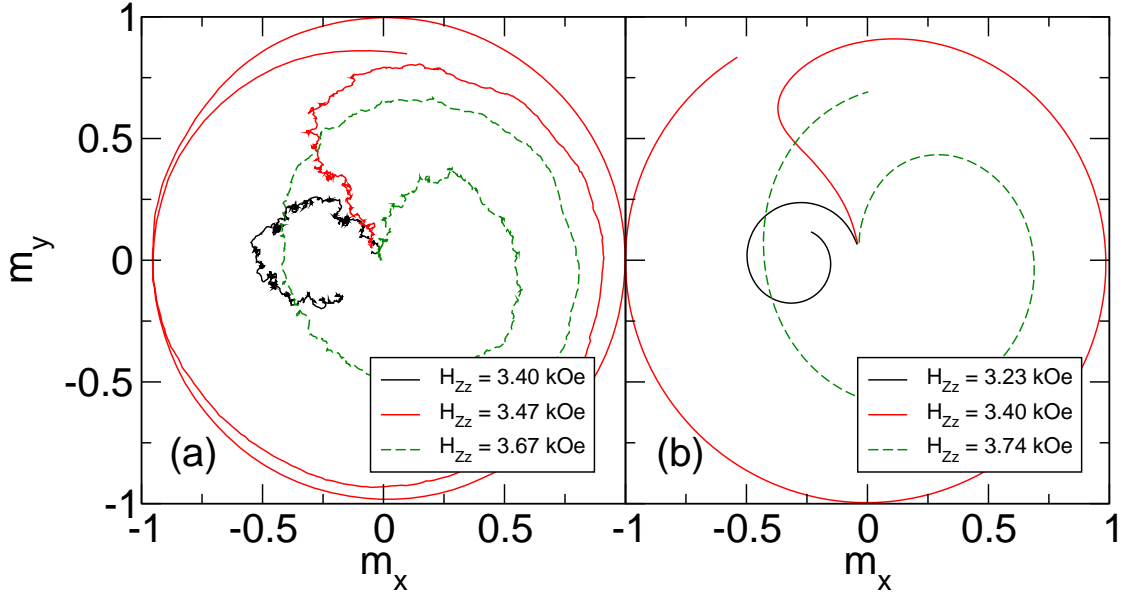}
\caption[]
{
(Color online) Paths of the end spin projected onto the $m_x$-$m_y$ plane for several values of $H_{\mathrm{Z}z}$.
The figure on the left (a) depicts the paths for $H_{\mathrm{Z}x} = 68$~Oe at $T=20.2$~K for times $0\leq{t}\leq2.67$~ns.
The black path for $H_{\mathrm{Z}x} = 3.40$~kOe precesses around a relatively stationary local field during this time.
For larger applied fields, the paths rotate opposite to those subjected to $H_{\mathrm{Z}z} < 3.40$~kOe.
Reduced switching times are observed around $H_{\mathrm{Z}z} < 3.47$~kOe that result from the end spin mostly rotating towards $m_z=0$, instead of around the easy axis.
Also shown on the right (b) are similar trials for $T=0$~K, with $H_{\mathrm{Z}x} = 170$~Oe for the times $0\leq{t}\leq1.34$~ns.
We note that the behavior of the end spin is relatively deterministic, even at $T=20.2$~K, however the average switching time is increased compared to the $T=0$~K trials due to longer path lengths from the stochastic fluctuations.
}
\label{fig:mxmyreent}
\end{figure}

Since switching in the pillar simulations initiates at the ends, we investigate the dynamics of the end spins and discuss the differences to characterize the ``reentrant'' behavior.
Figure~\ref{fig:mxmyreent} depicts the trajectories of the end spin for several values of $H_{\mathrm{Z}z}$.
For Fig.~\ref{fig:mxmyreent}(a) ($T=20.2$~K, $H_{\mathrm{Z}x}=68$~Oe, $0\leq{t}\leq2.67$~ns), the trajectory for $H_{\mathrm{Z}z} = 3.40$~kOe is remarkably different than the others, which is the result of the end spin precessing around a relatively stationary local field during this time frame.
For stronger applied fields, the local field at $t=0$~ns is reduced, which allows the end spin to relax towards the global free-energy minimum.
This happens for $H_{\mathrm{Z}z} = 3.47$~kOe, with the end spin essentially rotating in the longitudinal direction toward $m_z = 0$ during the early part of the switching process, resulting in a shorter mean switching time.
However, a maximum mean switching time is observed near $H_{\mathrm{Z}z} = 3.67$~kOe that exhibits an end-spin (and local field) rotation opposite the spin precession at $H_{\mathrm{Z}z} < 3.40$~kOe in the transverse plane in addition to a slower rotation in the longitudinal direction.
At even larger applied fields the switching time is reduced, which is the result of a very small or negative ($z$-direction) local field at early times and a faster domain-wall propagation during switching.
Additionally, due to the effect of the noise on the trajectories, the $20.2$~K trials take longer to move around the $m_x$-$m_y$ plane, compared to the $T=0$~K trials seen in Fig.~\ref{fig:mxmyreent}(b), with $H_{\mathrm{Z}x} = 170$~Oe for the times $0\leq{t}\leq1.34$~ns.

\section{Low Resolution}
\label{LR}
Finally, the lowest-resolution model of the physical system is a single spin, $1\times1\times1$.
Anisotropy in the previous two models is provided through the dipolar field, which is absent in the single-spin model.
However, we can approximate the effects of these fields using a crystalline anisotropy field.
The magnitude of the components of the corresponding uniaxial anisotropy field are found by calculating the shape-induced anisotropy (SIA) derived from a pillar of the same dimensions that is uniformly magnetized parallel to its easy axis.~\cite{Aharoni:dffrfp, Rhodes:deoumrb}
In practice, this involves first finding the induced magnetic surface charge at the ends of the pillar due to the initial magnetization.
These surface charges in turn create a magnetic scalar potential that is used in the calculation of the magnetostatic self-energy.
  
Once an expression for the magnetostatic self-energy of the nanopillar is found, the magnetometric demagnetizing factor in the $z$-direction, $D_z$, is defined as the factor that makes the magnetostatic self energy per unit volume equal to $2{\pi}D_z{m_s}^2$.
For a cuboid with equal width and length such as our model, this can be reduced to Eq.~($5$) of Ref.~[\onlinecite{Aharoni:dffrfp}], which is the form used in this paper.
More details of this calculation are provided in the Appendix.
This unitless factor has the property that $D_x+D_y+D_z=1$, where $D_x$ and $D_y$ are the magnetometric demagnetizing factors in the directions transverse to the easy axis.
Following Aharoni's convention,~\cite{Aharoni:dffrfp} we find $D_x = D_y = 0.4846$ and $D_z = 0.0308$ for the shape-induced anisotropy for our model's dimensions. 
This result, however, overestimates the coercive field of the pillar since the endcap formation of the higher-resolution models lowers the free-energy barrier for switching and is not accounted for in the calculation of the SIA term.
Consequently, higher switching fields are observed for this model when compared to the higher-resolution models.

\begin{figure}[tb]
\centering\includegraphics[scale=0.40]{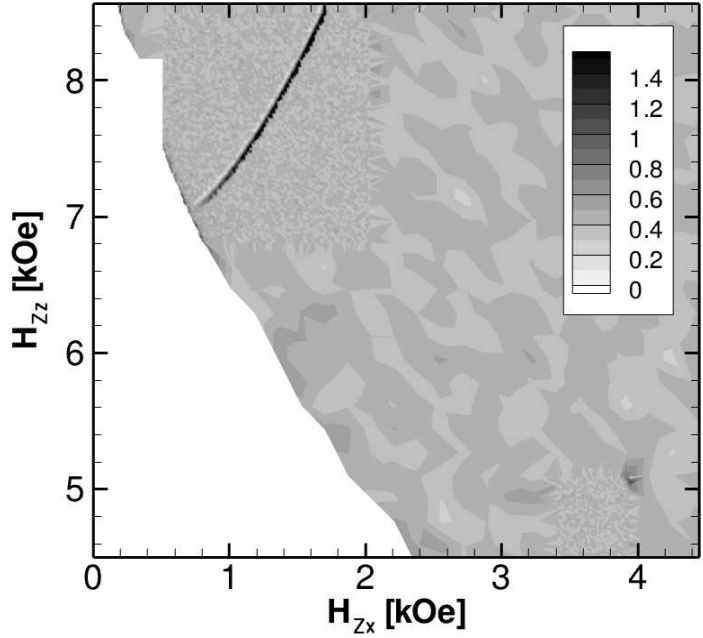}
\caption[]
{
Applied-field space at $20.2$~K for the lowest-resolution $1\times1\times1$ representation.
Using the same contour variable as before and a waiting time of $3.34$~ns, bimodal distributions are observed as an interior ridge that results from the precession behavior of the single spin near $M_z =0$.
The large white area in this figure indicates a region where switching times are greater than the waiting time.
}
\label{fig:switchspace_low}
\end{figure}

For this macrospin approximation, we observe bimodal behavior in the interior of the applied-field plot, shown in Fig.~\ref{fig:switchspace_low} using $C$ defined in Eq.~(\ref{eq:C}).
This region is now found as an internal ridge beginning at about $H_{\mathrm{Z}z} = 7.1$~kOe and extending up and to the right in the figure.
The location of this ridge is notably different than the one seen in the medium-resolution model, which relies on metastability that leads to bimodal behavior.
However, switching near the ridge in the macrospin model involves an applied field whose magnitude makes the magnetization dynamics essentially deterministic. 
Since the coarse-graining of the pillar is extreme in this case, the single-spin fluctuations resulting from the temperature are very small.
As it turns out, the bimodal ridge reveals a switching process that is sensitive to these tiny fluctuations, based on our definition of a switching event. 

\begin{figure}[tb]
\centering\includegraphics[scale=0.40]{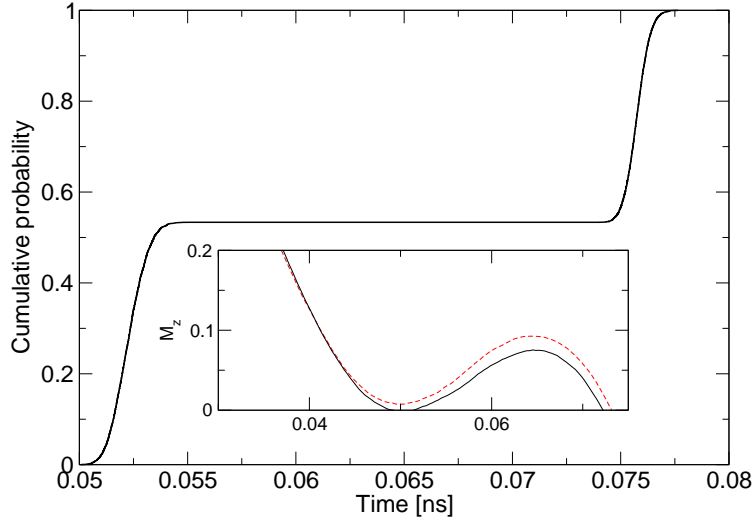}
\caption[]
{
(Color online) Cumulative switching-time distribution for $N=10000$ trials of the lowest-resolution pillar at $20.2$~K for the applied field $H_{\mathrm{Z}x} = 1.19$~kOe and $H_{\mathrm{Z}z} = 7.6$~Oe.
The inset depicts the $z$-component of the magnetization as a function of time and reveals a faster trial (black, solid) that crosses $M_z = 0$ on the first attempt due to a thermal fluctuation, while a slower trial (dashed, red online) requires an additional attempt to cross $M_z = 0$.
}
\label{fig:CD_1x1x1}
\end{figure}

A plot of the cumulative switching-time distribution is shown in Fig.~\ref{fig:CD_1x1x1} for a point in the bimodal region and reveals a clear separation of faster and slower switches with a relatively broad gap in time between the two switching regions, where no switches occur at all.
As with the medium-resolution model, the behavior of $M_z$ with time exposes the mechanism responsible for the bimodal distribution.
The precession of the single spin close to $M_z = 0$, the first crossing of which constitutes our definition of a switching event, leads to the observed distribution in the macrospin approximation.
The inset in Fig.~\ref{fig:CD_1x1x1} shows $M_z$ as a function of time for a faster switch (solid, black) and a slower switch (dashed, red online).
Faster switching is caused by the first precession becoming thermally ``knocked'' below $M_z = 0$, resulting in a shorter switching time.
Switches that do not cross $M_z = 0$ during the first attempt subsequently reach this value on the next precession, which leads to a longer switching time.
As with the medium-resolution pillar, the ratio of faster to slower switches changes as the bimodal ridge is traversed.
The bimodality for the single-spin model is thus simply a reflection of the inadequacy of the customary definition of the switching time as a first-passage time in this case.

\section{Conclusions}
\label{Con}
We have studied the switching statistics of a simulated magnetic nanopillar for three different resolutions of the computational lattice, looking for switching-time distributions that are bimodal.
The bimodal distributions result from processes that depend on the resolution of the computational lattice and the inherent fluctuations for each resolution studied here.

Limited by the computational time, we only investigate the distribution for a single value of the applied field in the highest-resolution model near the coercive field.
The mechanism responsible for the observed bimodal switching-time distribution in this realistic model is revealed as a consequence of fluctuations that determine a switching trajectory which may or may not carry the system through a long-lived metastable configuration.
A more detailed study of the switching dynamics of this model will be published elsewhere.~\cite{Thompson:inPress}

For the medium and lower-resolution models, the much smaller computational time enables a full exploration of the applied-field space.
We find very different mechanisms leading to bimodal distributions for the switching times in these two lower-resolution models.

The medium-resolution model displays a bimodal distribution near the coercive edge of the applied-field space that depends on the timing of the release of the two endcaps.
Fluctuations in this model serve to help the end spins of the pillar overcome a free-energy barrier separating metastable and stable orientations of the magnetization.  
If the fluctuations result in a switch with only one endcap releasing, a longer average switching time occurs.
However, when both endcaps release approximately simultaneously, the average switching time is measurably shorter.
Both of these situations are present near the coercive edge in the medium-resolution pillar and are responsible for the observed bimodal distribution.
In addition, this model also has the best agreement with the experimental data of the real system.
This may indicate that the real system's metallic iron core has a smaller width than originally reported.

The medium-resolution model also exhibits reentrant behavior for applied fields that are moderately aligned with the easy axis.
Starting near the coercive edge, the mean switching times increase for larger applied fields.
This is due to the trajectory of the end spin during the early times of the switching event.
The fastest mean switching times occur due to both the end spin and its local field rotating in the longitudinal direction toward the global free-energy minimum immediately after the field reversal.
Slightly higher values of the applied field reduce the magnitude of the end spin's local field and result in a spin rotation in the transverse plane in addition to a slower rotation in the longitudinal direction.

Finally, for the lowest-resolution representation of the physical nanopillar as a single effective spin, a bimodal distribution is seen as a ridge that stretches across the interior of the applied-field space, away from the coercive edge.
The bimodal distribution in this model is a result of a precession that can pass through the magnetization value defining a switching event earlier or later, depending on the small thermal fluctuations.
Trials that do not pass this magnetization value early will consequently cross it during the next precession, resulting in the observed bimodal distribution.

For the three models studied in this paper, only the highest-resolution model adequately captures fluctuations that result in multiple switching paths in the free energy that may occur in real pillars of width larger than the exchange length.
Consequently, our results show that conclusions about physical processes in simulated systems must take into account the degree to which the model resolution can reflect the length scales of the physically relevant fluctuations.

\section*{Acknowledgments}

We would like to thank Yongqing Li for providing us with the experimental data used in Fig.~\ref{fig:HswTheta} and Stephan von Moln\'{a}r for helpful discussions.
This work is supported in part by NSF grant No. DMR-0444051. 
Computational work was performed at the Florida State University High Performance Computing Center.

\section*{APPENDIX}
\label{app}

We can approximate the shape-induced anisotropy for the macrospin model by evaluating the magnetostatic self-energy of a pillar with the same dimensions as the full model ($1\times1\times15$).
To simplify the calculation, the magnetization of the pillar is assumed to be uniform and parallel with the longest axis. 
The general expression of the magnetic scalar potential,~\cite{jackson:ce}
\begin{equation}
\Phi_M(\vec{r})=- \int_{V} \frac{\vec{\nabla}^\prime\cdot\vec{M}(\vec{r}^{\, \prime})} {\mid\vec{r}-\vec{r}^{\, \prime}\mid}dV^\prime+
\oint_{S}\frac{\vec{n}^\prime\cdot\vec{M}(\vec{r}^{\, \prime})}{\mid\vec{r}-\vec{r}^{\, \prime}\mid}dA^\prime
\label{eq:a1}
\end{equation}
will consequently drop the first integral on the right-hand side since $\vec{\nabla}\cdot\vec{M}(\vec{r})=0$.
The remaining surface integral, involving the effective surface charges $\vec{n}\cdot\vec{M}(\vec{r})=\pm\sigma=\pm m_s$ at the ends of the pillar, is used to calculate the mutual self-energy of the two end faces with dimensions $a\times b$, separated by a distance $c$, and the self-energy of each individual face by letting $c=0$.
Together, these three terms are all that is needed to approximate the self-energy of our model. 

For the mutual energy $E_\mathrm{mutual}$ of the two faces at the ends of the pillar, the magnetic scalar potential should be integrated across both surfaces such that
\begin{equation}
E_{mutual}= \int_{0}^{b} \int_{0}^{a} dx_1 dy_1 \sigma_1 \int_{0}^{b} 
\int_{0}^{a} dx_2 dy_2 \sigma_2 \frac{1}{\{(x_2 - x_1)^2 + (y_2 - y_1)^2 + c^2 \}^{1/2} }.
\label{eq:a2}
\end{equation}
Solution of this integral is straightforward, although tedious, and can be expressed as a single function,~\cite{Rhodes:deoumrb}
\begin{eqnarray}
F(p,q) &=& (p^2-q^2){\Theta}\left\{\frac{1}{(p^2+q^2)^{1/2}}\right\}+p(1-q^2){\Theta}\left\{\frac{p}{(1+q^2)^{1/2}}\right\} \nonumber \\ 
&+& pq^2{\Theta}\left(\frac{p}{q}\right)+q^2{\Theta}\left(\frac{1}{q}\right)+2pq \mathrm{tan}^{-1}\left\{\frac{q(1+p^2+q^2)^{1/2}}{p}\right\} \nonumber \\ 
&-& \pi pq-\frac{1}{3}(1+p^2-2q^2)(1+p^2+q^2)^{1/2}+\frac{1}{3}(1-2q^2)(1+q^2)^{1/2} \nonumber \\ 
&+& \frac{1}{3}(p^2-2q^2)(p^2+q^2)^{1/2}+\frac{2}{3}q^3,
\label{eq:a3}
\end{eqnarray}
where $p=b/a$, $q=c/a$, and $\Theta(x) = \mathrm{sinh}^{-1}(x)=\ln\{x+(1+x^2)^{1/2}\}$.
The mutual energy of the two end faces of the pillar is then given by
\begin{equation}
E_\mathrm{mutual} = 2a^3 \sigma_1\sigma_2 F(1,q),
\label{eq:a4}
\end{equation}
while the self-energy $E_\mathrm{self}$ of each face is
\begin{equation}
E_\mathrm{self} = a^3 {\sigma_{1,2}}^2 F(1,0).
\label{eq:a5}
\end{equation}
Hence, the total demagnetizing energy $E_D$ of the cuboid is
\begin{eqnarray}
E_D &=& 2E_\mathrm{self} + E_\mathrm{mutual} \nonumber \\ 
&=& 2a^3 {m_s}^2 [ F(1,0) - F(1,q) ].
\end{eqnarray}
The final step to this approximation involves the definition of the magnetometric demagnetizing factor in the $z$-direction $D_z$, which in our case has the definition~\cite{Aharoni:dffrfp}
\begin{equation}
D_z = \frac{E_D}{2\pi V {m_s}^2}.
\end{equation}
The remaining factors in the $x$ and $y$-direction, $D_x$ and $D_y$ are evaluated by noting that $D_x + D_y + D_z = 1$ and $D_x=D_y$.

\end{document}